# Towards "*Pt-free*" Anion-Exchange Membrane Fuel Cells: Fe-Sn Carbon Nitride-Graphene "*Core-Shell*" Electrocatalysts for the Oxygen Reduction Reaction

Enrico Negro[(a,b,c)], Antoine Bach Delpeuch[(a)], Keti Vezzù[(a,c)], Graeme Nawn[(a)], Federico Bertasi[(a,b)], Alberto Ansaldo[(d)], Vittorio Pellegrini[(d)], Beata Dembinska[(e)], Sylwia Zoladek[(e)], Krzysztof Miecznikowski[(e)], Iwona A. Rutkowska[(e)], Magdalena Skunik[(e)], Pawel J. Kulesza[(e)], Francesco Bonaccorso[(b)]*, Vito Di Noto[(a,c,f)]*

[(a)] Section of "Chemistry for the Technology" (ChemTec), Department of Industrial Engineering, University of Padova, in the Department of Chemical Sciences, Via Marzolo 1, I-35131 Padova (PD), Italy.

[(b)] Centro Studi di Economia e Tecnica dell'Energia «Giorgio Levi Cases», Via Marzolo 9, 35131 Padova (PD) Italy.

[(c)] Consorzio Interuniversitario per la Scienza e la Tecnologia dei Materiali (INSTM).

[(d)] Istituto Italiano di Tecnologia, Graphene Labs, Via Morego 30, 16163 Genova, Italy.

[(e)] Faculty of Chemistry, University of Warsaw, Pastuera 1, 02-093 Warsaw, Poland.

[(f)] CNR-ICMATE, Via Marzolo 1, I-35131 Padova (PD), Italy.

**ABSTRACT:** We report on the development of two new *Pt-free* electrocatalysts (ECs) for the oxygen reduction reaction (ORR) based on graphene nanoplatelets (GNPs). We designed the ECs with a *core-shell* morphology, where a GNP *core* support is covered by a carbon nitride (CN) *shell*. The proposed ECs present ORR active sites that are not associated to nanoparticles of metal/alloy/oxide, but are instead based on Fe and Sn sub-nanometric clusters bound in *coordination nests* formed by carbon and nitrogen ligands of the CN *shell*. The performance and reaction mechanism of the ECs in the ORR are evaluated in an alkaline medium by cyclic voltammetry with the thin-film rotating ring-disk approach and confirmed by measurements on gas-diffusion electrodes. The proposed GNP-supported ECs present an ORR overpotential of only *ca.* 70 mV higher with respect to a conventional Pt/C reference EC including a XC-72R carbon black support. These results make the reported ECs very promising for application in anion-exchange membrane fuel cells. Moreover, our methodology provides an example of a general synthesis protocol for the development of new *Pt-free* ECs for the ORR having ample room for further performance improvement beyond the state of the art.

## ■ INTRODUCTION

The development of innovative, highly efficient and environmentally friendly energy conversion devices is one of the major challenges faced today by both fundamental and applied research.[1] In this respect, fuel cells (FCs) show great promise owing to their high energy conversion efficiency (values of up to 60% or more are currently demonstrated)[2] and negligible emissions of pollutants and greenhouse gases.[3] Fuel cells based on polymeric ion-exchange membranes and operating at low temperatures, T < 250°C, (*e.g.*, proton-exchange membrane fuel cells, PEMFCs, and high-temperature proton-exchange membrane fuel cells, HT-PEMFCs), attract considerable attention due to their very high power density (on the order of several hundreds of mW·cm$^{-2}$ or more [4-6]) and easy construction in comparison with other types of FCs (*e.g.*, molten carbonate fuel cells, MCFCs).[7-9] As an example, MCFCs require complex and bulky ancillary systems to recycle the $CO_2$ produced during operation; on the other hand, such systems are not needed in FCs based on polymeric ion-exchange membranes.[7-9] Massive research efforts have been dedicated to FCs,[10-12] with a particular reference to those based on polymeric ion-exchange membranes and operating at low temperatures. These systems typically adopt acid electrolytes (*e.g.*, perfluorinated proton-exchange ionomers, or polybenzimidazole imbibed with phosphoric acid).[8,13,14] Consequently, one of the critical bottlenecks in the operation of these systems remains the sluggish kinetics of the oxygen reduction reaction (ORR) process at the cathode.[7,15-17] Here, the currently used cathodic electrocatalysts (ECs) are based on platinum-group metals (PGMs) for optimal operation in an acid medium, since *Pt-free* ECs still yield low performances in these conditions.[15,18] The abundance of PGMs in the Earth's crust is very low, on the order of 0.01 ppm.[19] Furthermore, the geographical distribution of PGM resources is very uneven, raising significant risks of supply bottlenecks.[20] This hampers the widespread rollout of these low-T FCs.[21] Recently, the aforementioned issues have been addressed by the development of anion-exchange membrane fuel cells (AEMFCs).[22,23] In these devices the ORR takes place in an alkaline environment; accordingly, it occurs with a very different reaction mechanism in comparison with the acid medium.[24] Briefly, in the alkaline environment the first adsorption of dioxygen and the first electron transfers are facilitated if compared with the same processes in the acid medium.[25] Consequently, in the alkaline medium the ORR kinetics on PGM-free ECs is faster.[25] As of today, the performance level of PEMFCs is compatible with the standards required for practical applications;[26] typical power and current density values are on the order of ~700 mW·cm$^{-2}$ and ~2000 mA·cm$^{-2}$, respectively.[6,27] Recently, AEMFCs mounting Pt-based ORR ECs and reaching a performance level comparable to the one of PEMFCs have been demonstrated.[28] Accordingly, it can be inferred that state-of-the-art hydroxide-exchange membranes yield an ionic conductivity comparable to the one shown by the typical proton-exchange membranes adopted in PEMFCs, and no longer bottleneck AEMFC operation.[22] Anion-exchange membrane fuel cells based on PGM-free ORR ECs are able to demonstrate power and current densities values on the order of ~50 mW·cm$^{-2}$ and ~200 mA·cm$^{-2}$, respectively.[29-31] Although these values are rather promising, they are still ~10-15 times lower in comparison with the one demonstrated by PGM-based ORR ECs. Thus, the development of advanced, highly performant PGM-free ORR ECs, coupled with a sufficient durability in an alkaline medium, is an objective of utmost importance [32] which is, however, far from being achieved.

Here, we design innovative *core-shell*, *Pt-free* carbon nitride (CN) ORR ECs based on Earth-abundant metals, which are anchored by covalent metal-carbon and metal-nitrogen bonds on the support. The proposed ECs consist of graphene nanoplatelets (GNPs) *cores* covered by a thin CN matrix *shell* embedding active metal complexes sites. Graphene nanoplatelet *cores* are adopted with the aim to exploit: (a) their negligible microporosity, to facilitate the mass transport of reactants (*e.g.*, oxygen) and products (*e.g.*, water) in the cathodic electrocatalytic layer of the AEMFC; and (b) their high electrical conductivity, to minimize the ohmic losses associated with the electron transport from the active sites to the external circuit.[33-37] The proposed graphene-based ECs are significantly different than those described in the literature. In fact, the latter typically include oxide or nitride nanoparticles bearing the active sites to promote the ORR in an alkaline environment.[38,39] Furthermore, the ECs described in literature do not comprise a CN matrix *shell*.[40,41] Differently, in the proposed *core-shell Pt-free* ECs, the CN *shell*: (i) is templated on the GNP *core*; and (ii) coordinates the ORR active sites based on iron and tin. Thus, in the proposed *core-shell Pt-free* ECs the active sites are not found on the surface of inorganic nanoparticles forming a distinct component within the system. Iron is chosen for two reasons: (i) it gives rise to active sites where the first adsorption of dioxygen is facilitated; this affords fast ORR kinetics;[16,24] and (ii) it easily forms strong metal-carbon coordination bonds with the CN matrix (the *shell*), improving the stability of the ORR active sites.[42] We propose the use of tin as a co-catalyst in the ORR process owing to: (i) its high stability in the Sn-C bonds, which,[43] in a synergic way, stabilize the iron species in the active sites; (ii) its amphoteric character,[42] which facilitates the adsorption of dioxygen in the alkaline environment. The two ECs here proposed are labeled as FeSn$_{0.5}$-CN$_l$ 900/GNP and FeSn$_{0.5}$-CN$_l$ 900/GNP$_A$ in agreement with the widely accepted nomenclature.[16]

## ■ EXPERIMENTAL SECTION

**Reagents.** Dimethyltin dichloride, 95% and potassium hexacyanoferrate (II) trihydrate, 98% are purchased from ABCR and Sigma-Aldrich, respectively. Sucrose, molecular biology grade is obtained from Alfa Aesar. Graphene nanoplatelets are purchased from ACS Material, LLC. Potassium hydroxide (KOH, 98.4 wt%), hydrofluoric acid (HF, 48 wt%) and perchloric acid (HClO$_4$, 67-72%) are bought from VWR International, Sigma-Aldrich and Fluka Analytical, respectively. Isopropanol (> 99.8 wt%) and methyl alcohol (> 99.8 wt%) are purchased from Sigma-Aldrich. EC-10 electrocatalyst is acquired from Electro-Chem, Inc.; it has a nominal Pt loading of 10 wt% and is labelled *"Pt/C ref."* in the text. For reference gas-diffusion electrode (GDE) experiments, Pt(20%)/C from E-Tek is used. Vulcan® XC-72R is supplied by Carbocrom s.r.l.; it is washed with H$_2$O$_2$ (10 vol.%) prior to use. All the reactants and solvents are used as received and do not undergo any additional purification procedure. Doubly distilled water is used in all the experiments.

**Synthesis of the electrocatalysts.** [16,44] 1 g of GNPs, 1 g of sucrose and 2 mL of methanol are ground together for 9 hours in an agate jar using a Retsch PM100 ball mill. The resulting product is treated with HF, 10 wt% for 2 hours and finally rinsed extensively with water, yielding the support *GNP*. 355 mg of sucrose are dissolved in the minimum amount of methanol (*i.e.*, ~ 40 mL); 355 mg of GNPs are subsequently added to the solution; the resulting dispersion is: (i) homogenized for 2 min with a Bandelin Sonoplus HD 2200 probe sonicator, using ultrasonic impulses lasting 0.3 sec each and separated by pauses 0.7 sec long at a power rating of 20%; and (ii) transferred into a Teflon® beaker. The dispersion is heated at ~60°C in an oil bath and brought to a small volume (~1 mL) under vigorous stirring; the product is allowed to cool to room temperature. 141 mg of potassium hexacyanoferrate (II) trihydrate are dissolved into the minimum amount of water (~1 mL); the resulting solution is added to the dispersion comprising the GNPs, which is then homogenized extensively with a probe sonicator adopting the same experimental parameters previously described. 37 mg of dimethyltin dichloride are dissolved into the minimum amount of water (~1 mL), yielding a solution which is finally added to the product above; the resulting dispersion is homogenized extensively with a probe sonicator (same parameters as above), vigorously stirred for a few minutes, allowed to rest for 24 hours and finally dried in an oven for 120°C. The product of this process is placed into a quartz tube, where it undergoes a pyrolysis process comprising three steps, as follows: Step 1: 150°C, 7 hours; Step 2: 300°C, 2 hours; Step III: 900°C, 2 hours. The entire pyrolysis process is carried out under a dynamic vacuum of ~1 mbar. The product of the pyrolysis process is split into two aliquots. The first is treated three times with water, yielding the electrocatalyst labeled *"FeSn$_{0.5}$-CN$_l$ 900/GNP"*. The second aliquot is treated with HF 10 wt% for two hours, thoroughly washed with water and finally transferred into a quartz tube, where a second 2h pyrolysis step is carried out under a dynamic vacuum of ~1 mbar at 900°C (activation process, "A"). This latter process yields the electrocatalyst indicated as *"FeSn$_{0.5}$-CN$_l$ 900/GNP$_A$"*.

**Instruments and methods.** The wt% of C, H and N in the ECs is obtained by elemental analysis using a FISONS EA-1108 CHNS-O system. The assay of K, Fe and Sn is evaluated by Inductively-Coupled Plasma Atomic Emission Spectroscopy (ICP-AES). These measurements are carried out with a

SPECTRO Acros spectrometer with EndOnPlasma torch. The digestion of the samples is performed as described elsewhere.[45] The only difference, with respect to the published procedure,[45] relies on the fact that the initial oxidation step is carried out at 850°C for two hours in a ventilated oven, instead of 600°C for three hours.[46] The emission lines are: λ (Fe) = 259.940 nm, λ (K) = 766.490 nm, λ (Sn) = 189.926 nm. Thermogravimetric studies at high resolution are carried out between 30 and 1000°C by means of a TGA 295 analyzer (TA instruments). The instrument sensitivity ranges from 0.1 to 2%·min$^{-1}$; the resolution is 1 μg. The heating ramp is adjusted on the basis of the first derivative of the weight loss, and can range from 50 to 0.001°C·min$^{-1}$. The measurements are performed with an open Pt pan in an oxidizing atmosphere of dry air. Powder X-ray patterns are measured using an eXplorer diffractometer (GNR Instruments) mounting a monochromatized Cu K$_α$ source in the 2θ range 10-70° with a 0.05° step and an integration time of 40 sec. The MAUD software is used to analyze the data.[47] Transmission electron microscopy studies, both conventional and at high resolution, are executed with a Jeol 3010 apparatus mounting a high-resolution pole pieces (0.17 nm point-to-point resolution) and a Gatan slowscan 794 CCD camera. All the samples are prepared in agreement with a protocol described elsewhere.[48]

**CV-TF-RRDE measurements.** The inks for the realization of the electrode are prepared as described elsewhere.[49] 15 μL of each ink are pipetted onto the glassy carbon (GC) disk of a rotating ring-disk electrode (RRDE) tip. The latter is spun at ~700 rpm in the open air during drying to ensure that the final electrode film is uniform.[50] The loading of both FeSn$_{0.5}$-CN$_I$ 900/GNP and FeSn$_{0.5}$-CN$_I$ 900/GNP$_A$ on the GC disk of the RRDE tip is equal to 0.765 mg·cm$^{-2}$; only in the case of the Pt/C ref., the Pt loading on the RRDE tip is 15 μg$_{Pt}$·cm$^{-2}$. The details of the electrochemical instrumentation and setup used to carry out the measurements are reported elsewhere.[46] Briefly, the RRDE used as the working electrode is mounted on a Model 636 rotator (Pine Research Instrumentations); the collection efficiency of the Pt ring is equal to 0.39. The experiments are carried out with a Bio-Logic VSP multi-channel potentiostat/galvanostat. In a first step, the electrode is dipped in an acid 0.1 M HClO$_4$ solution and cycled between $E$ = 0.05 and 1.05 V vs. RHE under pure O$_2$ at a sweep rate of 100 mV·s$^{-1}$ as the RRDE tip is rotated at 1600 rpm. Cycling is interrupted as the voltammogramms become stable. The same electrode is subsequently dipped in a 0.1 M KOH solution and cycled as described above between $E$ = 0.05 and 1.05 V vs. RHE until the voltammogramms become stable; the final data are collected at ν = 20 mV·s$^{-1}$ as the rotator is spun at 1600 rpm. The ring electrode is kept at $E$ = 1.2 V vs. RHE for the detection of H$_2$O$_2$.[51] Hg/HgSO$_4$/K$_2$SO$_{4(sat.)}$ and Hg/HgO/KOH$_{(aq)}$(0.1 M) reference electrodes are used for the acid and alkaline environments and are placed in a separate compartment during the measurements. The potential is reported in terms of the reversible hydrogen electrode (RHE) scale, calibrated before each experiment in agreement with the procedure described in the literature.[52] The faradic ORR currents are obtained by subtracting from the voltammogramms described above other voltammogramms collected exactly in the same conditions, but after saturating the KOH solution with N$_2$.[53] iR-correction is carried out adopting a procedure described in the literature.[54] The gases used to saturate the electrochemical cell (i.e., high-purity oxygen and nitrogen) are obtained from Air Liquide. The disk currents are normalized on the geometric area of the GC.

**Galvano-dynamic measurements.** The electrochemical activity of the ECs has also been tested with a home-made gas-diffusion electrode (GDE, geometric area of the active part, 0.916 cm$^2$) mounted into a Teflon holder (with provision for gas feeding from the back of the electrode) and containing Pt ring as a current collector. The experiments have been performed with CH Instruments (Austin, TX, USA) Model 920D workstation. A saturated calomel electrode is used as a reference electrode and all potentials are expressed against the reversible hydrogen electrode (RHE). Platinum sheet serves as a counter electrode. The gas diffusion backing layer for the ECs is a carbon cloth (Designation B, 30% Wet Proofing; BASF Fuel Cell Co). Inks have been prepared by grinding in an agate mortar the ECs materials together with Vulcan® XC-72R (Cabot, USA) at the 1 to 1 mass ratio, and 2–propanol (POCh, Poland) (1 mL per 10 mg of the catalyst + Vulcan) and 5% Nafion®-1100 resin solution (Sigma-Aldrich). Final content of Nafion is 10% relative to the weight of the EC + Vulcan. In the fabrication of GDEs, the mass ratios of Pt in the Pt/C ref. vs. the proposed ECs is the same as in the CV-TF-RRDE approach, namely on the level of 1 to 51. After sonication for 1 h, followed by mixing under magnetic stirring for 1 h, the appropriate volumes of inks have been dropped onto the carbon cloth, placed onto a hotplate warmed up to the temperature of 353 K. Catalytic layers have been dried at 373 K to a constant weight followed by pressing them upon application of 2 kg·cm$^{-2}$ for 30 s. The loading of the ECs is ~2 mg·cm$^{-2}$. The measured currents have been recalculated against exact masses of each catalyst (expressed as specific currents). Pre-conditioning protocol has been identical to that for CV-TF-RRDE measurements. After open circuit potential (OCP) stabilization in oxygen-saturated 1 M KOH, galvanodynamic curves have been collected at the current scan rate of 1 mA·s$^{-1}$ and the oxygen flow of 50 mL·min$^{-1}$.

## ■ RESULTS AND DISCUSSION

**Figure 1(a)** shows the schematic of the experimental procedure adopted for the realization of the proposed ECs, i.e., by using GNPs covered with a tin and iron Z-IOPE (zeolitic-inorganic organic polymer electrolyte) precursor. The latter is produced by using a sucrose binder, followed by a multi-step pyrolysis process.[25,44,55] It should be observed that FeSn$_{0.5}$-CN$_I$ 900/GNP$_A$ EC is obtained by carrying out an activation process ("A") on FeSn$_{0.5}$-CN$_I$ 900/GNP as described in Section 2.2.

The chemical composition of the ECs is obtained by ICP-AES and elemental analysis; it is shown in Table S1 of Supporting Information and highlights that:

(i) the hydrogen concentration in FeSn$_{0.5}$-CN$_I$ 900/GNP is lower than 0.2 wt%, witnessing a very high graphitization degree of the CN *shell*;[16]

(ii) the wt% of nitrogen in the CN *shell* is lower than ~3 wt%. Indeed, nitrogen is only found in the CN *shell*, which comprises ~10 wt% of the whole EC (see Figure 1(b)). Thus, the concentration of nitrogen in the CN *shell* is ~10 times that determined for the whole ECs with microanalysis. For this reason, since the nitrogen concentration is at most 0.29 wt% (see Table S1 of Supporting Information), we can deduce that the maximum weight% of nitrogen in the *shell* is ~3 wt%. In these conditions, in agreement with literature results on similar CN systems, we can assume that the electrical conductivity of the CN matrix is high enough (> ~1-2·10$^{-2}$ S·cm$^{-1}$) to prevent significant ohmic losses.[16]

(iii) Both iron and tin are detected in FeSn$_{0.5}$-CN$_l$ 900/GNP; in fact, the tin/iron molar ratio is 0.1, which is quite different from the value expected on the basis of the reagents stoichiometry (*i.e.*, 0.5). This result indicates that tin is more easily eliminated than iron during the pyrolysis and washing processes performed during the preparation of FeSn$_{0.5}$-CN$_l$ 900/GNP. Finally, a nitrogen/iron molar ratio of 0.25 is indicative of a significant amount of iron atoms that are not directly involved in coordination by the nitrogen-ligand functionalities of the CN *shell*.

With respect to FeSn$_{0.5}$-CN$_l$ 900/GNP, in FeSn$_{0.5}$-CN$_l$ 900/GNP$_A$:

(i) the graphitization degree of the CN *shell* is enhanced, as witnessed by the significant reduction of the wt% of hydrogen (from 0.19 to 0.09 wt%) and nitrogen (from 0.29 to 0.04 wt%).[16] Unfortunately, a widely-accepted, *"quantitative"* definition of *"graphitization degree"* is not available. In general, the lower the amount of heteroatoms (*e.g.*, hydrogen, nitrogen, oxygen, sulfur, and others) in a carbon-based material, the higher the *"graphitization degree"*.

(ii) Tin- and iron-based species are decreased from 1.3 wt% to 0.4 wt% and from 4.68 to 0.14 wt%, respectively. Thus, it is concluded that the activation process affects significantly the chemical composition of the pristine FeSn$_{0.5}$-CN$_l$ 900/GNP.

After the activation process the tin/iron and nitrogen/iron molar ratios of FeSn$_{0.5}$-CN$_l$ 900/Gr$_A$ increase to 1.5 and 1.13, respectively (see Table S1 of Supporting Information), revealing that:

(i) ~97% of iron species, which are not bound by carbon and nitrogen coordination functionalities forming the *coordination nests* of the CN matrix (the *shell*), are eliminated;

(ii) ~25% of tin species are removed. Thus, with respect to iron, the activation process has a lower impact on tin species. This fact could be due to the presence of Sn as Sn(IV) in FeSn$_{0.5}$-CN$_l$ 900/GNP$_A$. In fact, on the basis of the preparation method and results elsewhere reported[42] it is expected that the Sn species are presented as Sn(IV), which is the most stable state of tin in ambient conditions.

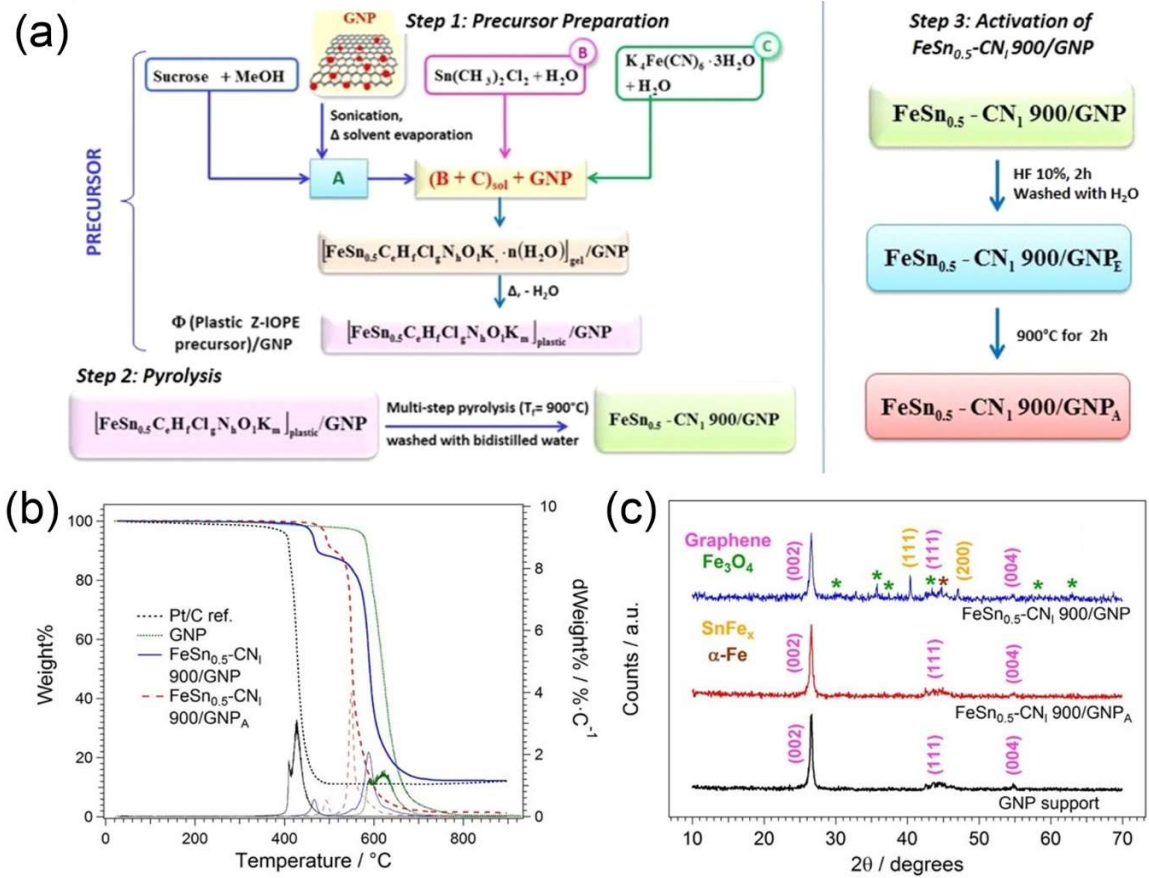

**Figure 1.** (a) Protocol for the preparation of the *core-shell* GNPs-based ECs. *FeSn$_{0.5}$-CN$_l$ 900/GNP$_E$* is the intermediate product of the activation process, after the etching with HF and before the final pyrolysis at 900°C; (b) HR-TGA profiles under an oxidizing atmosphere; (c) powder XRD patterns of ECs and GNP support.

The High-Resolution Thermogravimetric Analysis (HR-TGA) profiles in an oxidizing atmosphere of both FeSn$_{0.5}$-CN$_l$ 900/GNP and FeSn$_{0.5}$-CN$_l$ 900/GNP$_A$ ECs exhibit two clearly distinguished thermal degradation events: I and II (see Figure 1(b)). I occurs at $T_I$ < 500°C and is ascribed to the combustion of the CN *shell* in agreement with results determined on ECs that include a CN matrix with a similar nitrogen composition.[45] II, which takes place in the temperature range 540 < $T_{II}$ < 590°C, is associated to the oxidation of the remaining GNP *core*. Essentially, the activation process: (i) raises $T_I$, confirming previous results on the higher graphitization of the CN *shell*;[25] and (ii) lowers $T_{II}$, owing to the generation of an increased amount of defects in the GNP support *core*. The high-temperature residue of FeSn$_{0.5}$-CN$_l$ 900/GNP corresponds to $M_xO_y$ species with M = Fe, Sn.[25] The HR-TGA analysis demonstrates that FeSn$_{0.5}$-CN$_l$ 900/GNP$_A$ exhibits a negligible high-temperature residue (*ca.* 1.5%) (see Figure 1(b)) due to the low amount of metals in this EC (iron: 0.14%, tin: 0.4%; see Table S1 of Supporting Information). Therefore, with respect to the Pt/C reference (where $T_I$ occurs at ~426°C) the thermal stability of both pristine ($T_I$ ≈ 466°C) and activated ECs ($T_I$ ≈ 492°C) under oxidizing atmosphere is significantly improved. In fact, in state-of-the-art Pt/C ECs the thermal degradation under oxidizing atmosphere of carbon-based supports is promoted by the platinum nanoparticles.[56] On this basis, the proposed graphene-based platinum-free ECs are very stable likely for the lack of metal alloy nanoparticles that typically promote their thermal degradation in the presence of an oxidizing atmosphere.[56]

The powder X-ray patterns of the ECs, shown in Figure 1(c), exhibit: (i) a high-intensity and sharp peak at 2θ ~ 26.6°; (ii) a broad peak at 2θ ~ 44.5°; and (iii) a low-intensity peak at 2θ ~ 54.8°. These reflexes are attributed respectively to the (002), (111) and (004) peaks of the GNP component of the samples (COD #9008569), with the $P6_3/mc$ space group.[57] XRD results reveal that in the ECs, along the [001] direction the graphene layers are still partially stacked with a significant stacking fault disorder. This nano-morphology is probably modulated by the CN *shell* wrapping the graphene support. The X-ray pattern of FeSn$_{0.5}$-CN$_l$ 900/GNP also shows peaks associated to three additional phases, *i.e.*: (i) magnetite, Fe$_3$O$_4$, (COD#9006247);[57] (ii) alloy nanoparticles comprising a random solid solution of tin and iron, indicated as SnFe$_{0.39}$ (COD#9012707);[57] and (iii) α-Fe (COD#1100108).[57] The presence of Fe$_3$O$_4$, SnFe$_x$ and α-Fe phases, which belong respectively to the *Fd-3m*, *Fm-3m* and *Im-3m* space groups,[57] are in agreement with the chemical composition of the ECs summarized in Table S1 of Supporting Information. Structural information (*i.e.*, abundance, cell parameters and particle size) derived from the Rietveld analysis for FeSn$_{0.5}$-CN$_l$ 900/GNP is reported in Table S2 of Supporting Information. The X-ray pattern of FeSn$_{0.5}$-CN$_l$ 900/GNP$_A$ reveals only peaks associated to the GNP component, demonstrating that the activation process removes almost completely the metal-based nanoparticles incorporated in the pristine FeSn$_{0.5}$-CN$_l$ 900/GNP. The morphology of FeSn$_{0.5}$-CN$_l$ 900/GNP and FeSn$_{0.5}$-CN$_l$ 900/GNP$_A$ is shown in Figure 2.

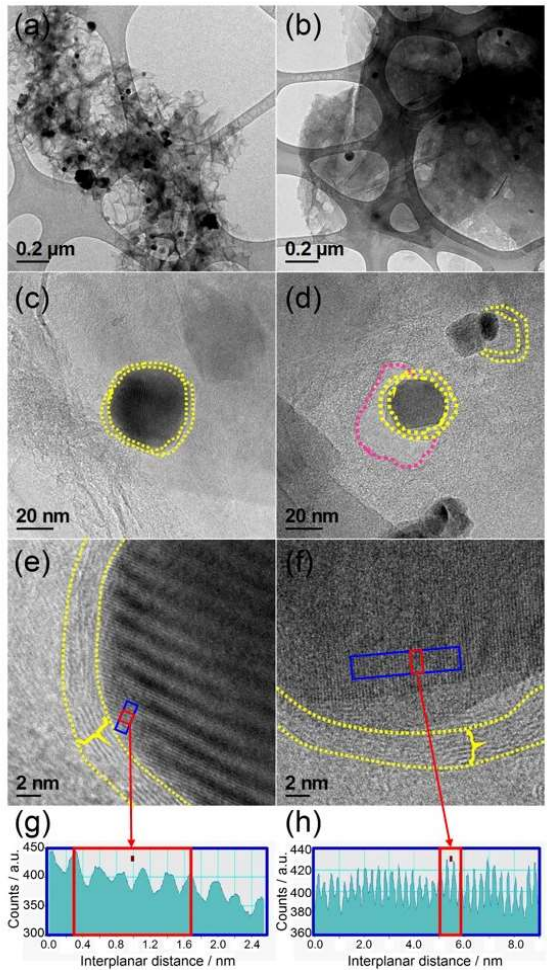

**Figure 2.** HR-TEM micrographs of: (a, c, e) FeSn$_{0.5}$-CN$_l$ 900/GNP; and (b, d, f) FeSn$_{0.5}$-CN$_l$ 900/GNP$_A$ at different magnifications. The inter-planar distances $d_{131}$ of the Fe$_3$O$_4$ phases of (g) FeSn$_{0.5}$-CN$_l$ 900/GNP and (h) FeSn$_{0.5}$-CN$_l$ 900/GNP$_A$ are shown.

FeSn$_{0.5}$-CN$_l$ 900/GNP shows small (d < 100 nm) nanoparticles (mostly based on Fe$_3$O$_4$), which are embedded in a CN *shell* covering the GNP *core*. In particular, the sub-100nm nanoparticles are encapsulated in a compact *onion-like* CN *shell* (highlighted in yellow in Figure 2(c) and 2(e)). The inter-planar distance evidenced in the small nanoparticles (Figure 2(e)) corresponds to the $d_{131}$ value of a typical Fe$_3$O$_4$ phase, confirming the identification of the latter in the powder X-ray pattern reported in Figure 1(c). After the activation process very few metal-based nanoparticles are observed in the *shell* of EC (see Figure 2(b)): the density of the metal-based nanoparticles in FeSn$_{0.5}$-CN$_l$ 900/GNP$_A$ is ~40-45 nanoparticles·μm$^{-2}$. The activation process of the ECs gives rise to a cratered CN *shell* covering the GNP *core* (*e.g.,* see the crater highlighted in purple in Figure 2(d)); the morphology of the CN *shell* is not altered. In few instances, the *onion-like* features of the CN shell matrix are still visible (see Figure 2(d) and 2(f)), and inhibit the etching of the sub-100nm nanoparticles underneath.

The ORR performance of the ECs in the alkaline environment is investigated by cyclic voltammogramms (see Figure 3(a)) and Tafel plots (see Figure 3(b)) [48].

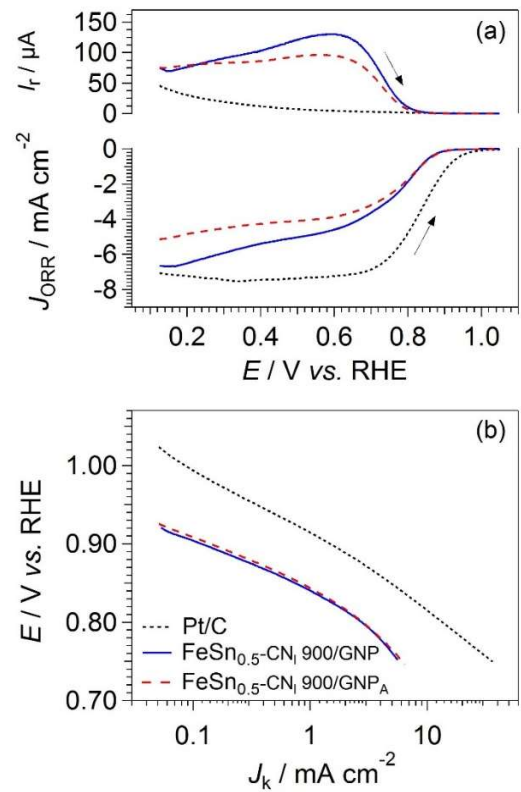

**Figure 3.** (a) CV-TF-RRDE anodic $J_{ORR}$ and $I_r$ profiles; and (b) corresponding Tafel plots of the ECs in O$_2$-saturated 0.1 M KOH; ω = 1600 rpm; ν = 20 mV s$^{-1}$; $T$ =298 K.

The ORR performance, selectivity and reaction mechanism of FeSn$_{0.5}$-CN$_l$ 900/GNP and FeSn$_{0.5}$-CN$_l$ 900/GNP$_A$ ECs are not influenced by the activation process. On these bases, it is inferred that: (i) the active ORR sites of both ECs are similar; (ii) the sub-100nm nanoparticles, that are mostly detected on FeSn$_{0.5}$-CN$_l$ 900/GNP only, play a negligible role in the ORR; and (iii) the ORR active sites are located on features present in both ECs, namely the iron and tin species stabilized in the *coordination nests* of the CN *shell*. With respect to the Pt/C reference: (i) the ORR overpotential of both the proposed ECs is only ~70 mV higher; (ii) the ORR mechanism is the same, as witnessed by the Tafel slope (see Figure 3(b));[24,58] and (iii) the selectivity determined as elsewhere reported[51] in the 4e-ORR pathway is lower, as indicated by the higher amount of H$_2$O$_2$ produced.

These results are rationalized admitting that during the ORR the first adsorption of dioxygen in the alkaline medium is followed by charge transfer occurring through an *outer-sphere* process[24]. A significant fraction of hydrogen peroxide, $X_{H2O2}$, is thus obtained (see Figure S3 of Supporting Information; $X_{H2O2}$ values on the order of 45% at $E$ = 0.5 V *vs.* RHE are detected), which is readily expelled from the electrode layer on the RRDE disk and eventually detected at the ring electrode. This evidence confirms that the proposed ECs exhibit a very low microporosity. In prospective this is a very useful feature, as it facilitates the mass transport of reactants and products. This mitigates the concentration overpotentials at high current densities in the FCs mounting on the cathode the proposed ECs.[59] Indeed, if the proposed ECs had a high microporosity, the hydrogen peroxide produced during the ORR would be trapped in the cavities of the EC, undergoing a further reduction to $H_2O$. This would allow extracting two more electrons from a single dioxygen molecule, boosting the energy conversion efficiency of the system at the expense of more severe concentration overpotentials in the final FC.

The galvano-dynamic profiles determined on GDEs (Figure 4) show that half-cell potentials yielded by the Pt/C reference are only 70-80 mV higher with respect to those obtained with the proposed ECs. Furthermore, the traces of $FeSn_{0.5}$-$CN_l$ 900/GNP and $FeSn_{0.5}$-$CN_l$ 900/$GNP_A$ ECs are almost coincident. In detail, the potential detected on the GDEs at a current density of 40 mA·cm$^{-2}$ is equal to 0.53 and 0.51 V vs. RHE for $FeSn_{0.5}$-$CN_l$ 900/GNP and $FeSn_{0.5}$-$CN_l$ 900/$GNP_A$ ECs, respectively. These results are in agreement with the outcome of CV-TF-RRDE experiments (see Figure 3(a)).

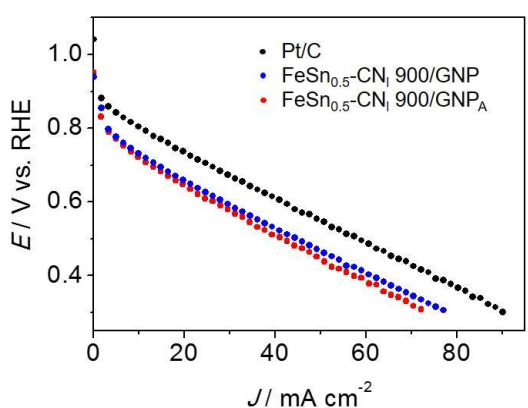

**Figure 4.** Galvano-dynamic steady-state polarization responses for ORR recorded using GDEs in $O_2$-saturated 1 M KOH solution; catalyst loading: ~2 mg cm$^{-2}$; $O_2$ flux = 50 mL· min$^{-1}$; $T$ = 298 K.

Accordingly, it is demonstrated that the proposed ECs are able to operate effectively both in almost ideal conditions (i.e., those achieved by the CV-TF-RRDE experimental setup[60]), and in an environment (i.e., a GDE) that closely mimics the cathodic electrode layer of a real AEMFC being affected by ohmic and concentration overpotentials. This result allows inferring that the proposed ECs have the potential to be successfully implemented in the fabrication of the cathodic electrocatalytic layer of an AEMFC. The development of ORR electrocatalysts able to operate without the introduction of parasitic overpotentials in the presence of methanol is an important objective of the research in the field of direct methanol fuel cells (DMFCs).[61,62] The tolerance of the proposed ECs towards methanol poisoning is gauged by recording the GDE polarization curves in electrolytes containing 1 M $CH_3OH$ (see Figure S4 of Supporting Information). Contrary to conventional Pt/C ECs reported in the literature,[63,64] the performance in the ORR of the proposed ECs is largely unchanged, thus implying tolerance of the proposed *Pt-free, core-shell* GNPs-based ECs to the presence of an organic fuel such as methanol.

## ■ CONCLUSIONS

Here, we design and synthesize *core-shell* ECs for the ORR process ($FeSn_{0.5}$-$CN_l$ 900/GNP), which includes metal-based nanoparticles embedded in a CN *shell* templated on graphene nanoplatelet *cores*. In addition we propose an activation process, which is suitable to improve the graphitization of the CN *shell* and removes most of the metal-based nanoparticles without a significant impact on the ORR performance, yielding the $FeSn_{0.5}$-$CN_l$ 900/$GNP_A$ sample. The overall multi-step synthetic approach followed to obtain the proposed ECs is very general and opens the door to a large number of further syntheses to obtain *Pt-free* ECs for the ORR with a well-controlled chemical composition, morphology and structure, capable of operation at overpotentials less than 100 mV higher with respect to the one of Pt and with a high selectivity towards the 4-electron reduction of dioxygen to water. Results reveal that the ORR active sites of the proposed ECs are: (i) not located on the metal-based NPs detected in pristine $FeSn_{0.5}$-$CN_l$ 900/GNP (i.e., $Fe_3O_4$, $SnFe_x$ and α-Fe); (ii) present as subnanometric metal clusters fixed on the surface of the CN *shell*, where they are stabilized by *coordination nests*. These latter are formed by aggregates of carbon and nitrogen ligands provided by the CN matrix. Accordingly, these subnanometric metal clusters are very stable. Indeed, they are not removed during the activation process and bestow to $FeSn_{0.5}$-$CN_l$ 900/GNP and $FeSn_{0.5}$-$CN_l$ 900/$GNP_A$ a remarkable ORR activity: indeed, the ORR overpotential of the proposed ECs is only ~70 mV higher than that of the Pt/C ref. despite the very low metal concentration (lower than 6 wt%) and total absence of PGMs. Finally, it is shown that the proposed ECs are: (i) highly tolerant to methanol poisoning; and (ii) able to effectively operate also in a gas-diffusion electrode setup, where the ORR performance is affected by charge and mass transport issues similar to those encountered at the cathodic electrocatalytic layer of a single AEMFC. This is a crucial results highlighting the feasibility of the proposed ECs to be implemented in a real device.

## ASSOCIATED CONTENT

Chemical composition of the ECs; structural information derived from the Rietveld analysis of $FeSn_{0.5}$-$CN_l$ 900/GNP; results of the powder XRD pattern decomposition by the Rietveld procedure of the $FeSn_{0.5}$-$CN_l$ 900/GNP EC; cyclic voltammogramms of $FeSn_{0.5}$-$CN_l$ 900/GNP, $FeSn_{0.5}$-$CN_l$ 900/$GNP_A$ and Pt/C ref. in $N_2$-saturated 0.1 M KOH; evolution of the selectivity towards production of $H_2O_2$ against the potential of the investigated ECs in $O_2$-saturated 0.1 M KOH; galvanodynamic steady-state polarization responses for ORR using GDE in $O_2$-saturated 1 M KOH and 1 M KOH + 1 M $CH_3OH$. This material is available free of charge via the Internet at http://pubs.acs.org.

## AUTHOR INFORMATION

### Corresponding Author

*Prof. Vito Di Noto; e-mail address: vito.dinoto@unipd.it

### Funding Sources

The research leading to these results has received funding from: (a) the European Commission through the Graphene Flagship – Core 1 project [Grant number GA-696656]; and (b) the Strategic Project of the University of Padova *"From Materials for Membrane-Electrode Assemblies to Electric Energy Conversion and Storage Devices – MAESTRA"* [protocol STPD11XNRY]. Partial support by Maestro Project [2012/04/A/ST4/00287 (National Science Center, Poland)] is also appreciated.

## ACKNOWLEDGMENT

We acknowledge Massimo Colombo and Mirko Prato for useful discussions.

SYNOPSIS TOC

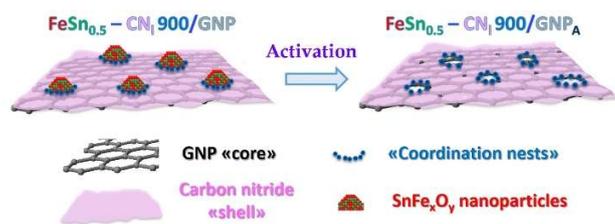